# Android Permission Model

Chalise Birendra,
Nanjing University of Science and Technology
biren884@hotmail.com.

*Abstract: The recent evolution on the smart phone technology has made its application market huge and less secure. Every single day large number of apps introduced in the android market (mostly on google play store) without any particular inspections which creates a lot of security issues and they remain unresolved. There are a lot of recent and increasing security issues which are mostly caused by the android apps.*

*Mainly, the access of user data through the apps relied on the user's given permission to apps. So, it's always a big question that how different apps on the android smart phones bypass the android permission model to have root access and how user data compromised or stolen by its apps. It's not startling that the Google Play Store has massive number of malicious apps that may gain users' attention to fall victim for one, but some time it might be even worse than we thought. In this paper, we will look at the security issues (i.e. rooting phone, malware bypassing and secondary storage) and how easy to get access to users phone by using its apps.*

*Furthermore, I will discuss the android update mechanism and its functionalities as well as vulnerability to possible third-party bypassing and root access. Moreover, a key feature and source of the all possible access route to android applications I will discuss an android permission model.*

*Finally, I will first time present concept of the feature based development model implementing a feature based permission selections as a solution to android application system and describe the purposed Feature based permission model in detail – how it is the solution for current security issues caused by intentional (evil coded) or unintentional (poorly coded) coding using android permission models.*

**Introduction**

Smart phones are very popular today and used for various purposes – from communicating around the world (personal or business purposes) to buying daily usage materials online on the internet, however most of the users are unaware of possible exploit i.e. how their information or confidentiality handled by developer [3] or vendors of the different application and its service providers. For smart phones and its apps, user is the most precious stakeholder but they are not taken in consideration while developing the applications.

Usually, Apps developed for users while permission models [1, 7] used by developers without securing user's information or privacy rights. Until now, user do not have much choice than accepting all the permission criteria set by developers even that is not necessary for apps or users. Production and distribution of poorly coded apps are all over the phone market today, which further exploits all the permissions to other apps that have evil codes runs in background to get the user information while they are acting as benign one.

Moreover, these kinds of apps find all over the current markets, which neither be detected as any threat by antivirus nor users can deny, skip or reject any unnecessary permission. Generally, malware uses these kinds of application to get the access over the users smart phones and compromise the information.

Similarly, most of the smart phone updates mechanism including android update mechanism works using third-party, which also allows malware and viruses [9] to bypass and gain the root access of the user's phone. So questions arises, is the permission model working for users? Are users making good decisions? The answer is "users are not making any decisions" – only developers and service providers are making decisions, which compromises the user's privacy and information. Overall, There are various permission for on the users phones such as edit SMS, read contact – mostly developer uses all of them than using required ones because if permission misses for particular feature then the application will not work. And for malware they take all information from regular poorly coded applications to prevent detection of anything suspicious activities by users.

A solution to all the security problems faced by users and developers is implementation of a feature based programming model, which helps both users and developers. It ensures protection of user's privacy as well as allows comprehensive solution to developers. Applications based on its feature, which will be a great achievement while it comes to complete implementation. Furthermore, this feature based model used on permission model of the android application for automatic generation of permission and manual user selections. It allows feature based selection process either by developer while developing or by users while using the application and its features. So, the concept has variety of solutions which offer to solve the problems. Now, we can include users to make decision for their information via permission models as well as improve the platform considering user concerns and make permission model independent and secure while developing apps.

**Related Works**
There is still a lot of work left to ensure full security of smartphones and its security systems. The Review on android and smart phone security, Research journal of computer and information technology science, the work done by Tiwari Mohini, Srivastava Ashish Kumar and Gupta Nitesh [12] shows the security concerns over android rather presenting any particular solution. Similarly, the work done by Vaibhav Rastogi, Yan Chen and Xuxian Jiang Evalutaing android malware against transform attack [9] is impressive toward malware system only. And the research on the mobile platform and solution and security solutions [13] in Samsung Corporation includes more about after install permission manager and tested on some of the phone.
However, this paper focuses on the all major security concerns of the users and the change of the apps permission model and purpose the developer and user-friendly permission model based on the feature of the android applications.

**Security issues in android apps**
The source of the problems in the android application system is either through misuse of the user given permissions or poor use of the permission models by the vendor. Here, we will discuss the main security issues (i.e. stealing and compromising user information) and its causes as well as general security measures to prevent possible causes.

*Rooting Android:* Usually rooting [13] done by different third-party software such as Vroot,SuperOneClick etc. using known exploit to the users i.e. GoogleBreak etc. –they can give you root so you can load custom firmware. There are a lot of Trojan Android apps on the official android market which pretends as a normal app, runs exploits and steals your data in the background. Root access to phone can edit, modify or delete the user security keys, erase the entire phone data, manage account tc. [11], which means basically root access gives permit to do almost everything on the smart phones.

*Malware Bypassing*: Malware [9] takes all the code (such as all permission for the user device) from the regular apps (i.e. games, chat, call or internet using apps) and calls it. So to the user it looks like the app is running normally but it runs evil code in the background. Generally, it runs the root exploits [9] if it is successful then it installs a system service and only firmware apps are left in the system.

However, System services have access to extra permissions and the permission model breaks down around system services. System app steals your data and sends it to a C&C. There are two different known root exploits for Android both were patched in Android 2.2.1 [3], which release was very earlier and after that some malware are breakout so for the time interval other malware can already be written. Malwares pretends as different apps like games, ads and dirty apps. Benign apps that have flaws that allows bad code to take advantage of it, and where information compromised without knowing by users. It's not about traditional attack vectors like buffer overflows to make apps not runnable or creating error it's become old and out of interest. What interested in here is getting permissions we shouldn't by piggybacking on poorly coded apps.

Secondary Storage (SD Card): Users' store their data in external storage, similarly third-party smart phone apps stores all data including sensitive data on the external storage cards aka SD card [6]. Unlike hard disks in computer systems usually SD cards are in VFAT format which is older not secure format for storing information. Moreover, everything that saved on SD cards - VFAT format is in simple document files and it can also be in human readable format while developing or saving through the application codes which is known as badly saved files. Here is code snippet from android programming as an example of badly saved code:

```
File root = Environment.getExternalStorageDirectory(); String filename ="IMEI"
```

The String file name IMEI is an example of badly coded and saved on the external storage directory of smart phone. When application installed, it create file on external storage with name of IMEI and saved the data in human or application readable format. So, hackers or malware take advantage of these badly saved files. Similarly, while developing file is use to send the information through application codes which is known as badly send files. Here is code snippet form android programming as an example of badly send code:

```
File String message = "IMEI: " + Line; String number = "15850779296";
```

Similarly, here the String file message IMEI and number are badly coded for message and send on the number in readable format. When applications interact with users and send the message of IMEI and number save the data in human or application readable format in the apps code can further be used by evil codes to read, write or modify the data. Malicious or Trojan apps discover how the data stored in the external storage i.e. either in human readable format or in encrypted and accesses it and sends it to an attacker. And the process is easier while apps are poorly coded and information store in easy or readable formats.

**General Security Measures**
Protection is always better than damage so, to prevent those security concerns developers and users need to consider different security measures or cautions to store information securely in smart phones.

*Not on SD Card:* The easier bypassing is cause of user permission to read data external storage so users' confidential information should not be stored in the external storage cards.

*Not in source code*: The confidential information about apps or user data like phone number, email, password etc. should not be stored in the application source codes.

*Not world readable:* Any kind of user data or confidential information should not be stored in the human readable format.

**Android Update Mechanism**
The substantial growth in a technological development and arising of new security concerns such as hacking, cracking, viruses and malware etc. – the constant update of the smart phone software is very important and almost inevitable.

For update android system [4] pushes over the air for android update mechanism [4, 14] (see figure 1.1, Android Update Mechanism), which is one of the best phone updating mechanism among all other smart phones update providers. However, the problem with Android is third-party, when Google releases new versions and Google phones get it but others have to port it. For example when you have to update Z4 or Sony companies android smart phones you should update from them than updating directly from the Google [14].

Usually, the instances of new versions pushing out to everyone take six or more months. Although, most of the old phones cannot be updated because of its hardware or software specification so now Google release development API level of 24 but developer should use API level of 8 to include all android user market.

Similarly, there are a lot of users who uses the different android version of 2, 3 and 4 while current Google release version is 5.1. So the third-party update and older phone which cannot update security patches, development error or new and constantly changing security mechanism further exploit user smart phones to various security loopholes i.e. malwares and makes access of user data easier. That means the update mechanism of the smartphones are also not completely secure because of the current third-party update mechanism over most of the phones [14].

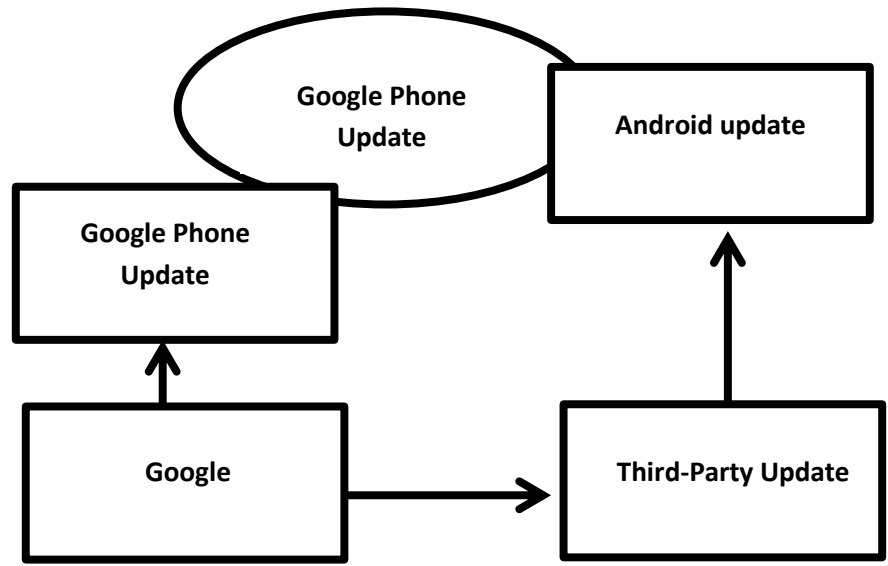

Figure 1.1: Android Update Mechanism

**Android Permission Models**
Android permission models [7] generally refer to the user provided or developer provided permission to application which is the main gateway between all the user information and vendor's applications. Though, the provided permission model completely ignores the user involvement over its system. In this section we will discuss the current model its vulnerabilities, security loop holes as well as the purposed solution and feature based models.

*Android Apps Permissions:* Permissions for accessing the phone is a key to all security problems over the smart phones software technologies. Without user is given, permissions neither any apps installed nor run, to install apps on smart phone so user must allow all necessary or unnecessary permission that apps want to use[1, 7]. There are various permission requires based on the app's features for correctly working of the apps. Usually, developers should use only required permissions but mostly they use all as piggy bag with or without knowing its disadvantages for their ease than selective as it's feature, which further leads to "steal of the user information" by evil apps such as malware using the permissions from the piggy bagged apps.

There are all types of permissions [7] which allow accessing every possible feature and information over the users' phones. They are as follows:
*Edit SMS, Read SMS, Send SMS, Receive SMS and Modify/delete:* These permissions use for reading, editing, sending, receiving, modifying and deleting from the phone by its applications.
*USB storage contents:* This permission use to read and write the USB storage and its contents.
*Prevent phone from sleeping:* This permission use to keep phone awake and prevent getting it locked. Read and write to your personal information including contact data.
*Phone calls, Services that cost you money:* These permissions use to read and write user contact and call information and alter it as well as cost user money.
*Act as an account authenticator:* By using this permission apps are able to edit and modify the related account information.
*Write sync settings GPS data:* This permission use to pinpoint location of the user's phone. Mostly it uses to display ads based on user locations and choices.
*Manage accounts:* This permission allows apps to edit and modify the user accounts. Read phone state and identity: This permission allows apps to read the current state of the phone and identify as its unique id.
*Full network access:* This permission allows full network access over user phones such as internet & its apps. So, apps use it for various features such as calls, sms, internet etc. which is user's worst nightmare, while enjoying the fastest growing technology users are also forced to compromise their information which leads to theft to more dangerous crimes. On android, permission for apps to use different phone state plays vital roles from loss of data to malware attacks.

**Android Developer Permission Model**

The Figure (Figure 1.2) shows Android permission model form android application system for developing apps. Developers normally use the official development platform [10] or java eclipse to write codes and run apps on phone, while doing so application will crashed over again and again even you rewrite code or change debugging statements until you have all the required permission. To prevent from crashing of applications the developers ask for all imaginable permissions at first even they are not using as features of the applications. Those unused permission are further used by malicious applications running in same or different system takes an advantage and gain the root access to steal user data. Malicious application finds those as easily hack/crack able apps to get user permission and information.

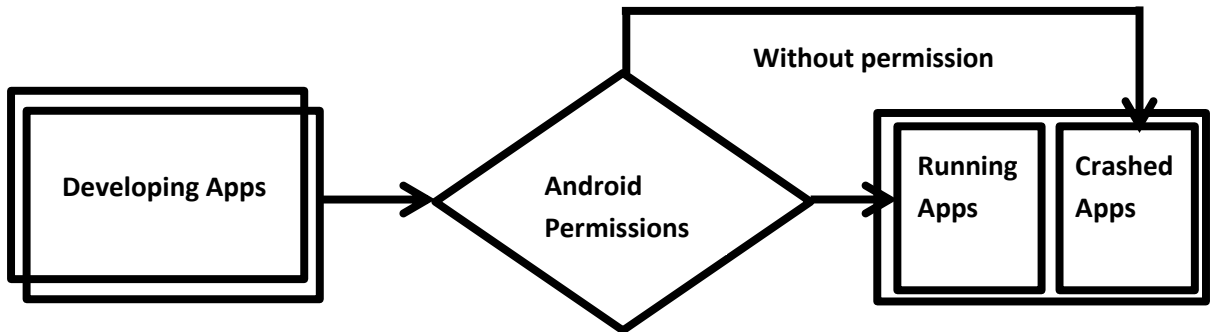

Figure1.2: Android developer permission model

The figure shows importance of the permission models for developer point of view as well as possible miss uses of from vendors or developers. Overall, the Trojan apps or malware apps uses other application with full user permission to gain the root access of the smart phones, while they are getting users' attention and acting as normal apps.

**Purposed Solution (Feature Oriented Development Models)**

The purposed solution is development of application based on its features [5] as well as involvement of the user as a key stakeholder of the model by giving them the choice of manual permission selection without any development or uses problems. Here I am presenting the possible solution as feature based permission model and manual permission selection procedure to the users.

**Feature based permission models**

The concept of feature based [5] permission models [7] for android is as an auto generated permission model (see figure 1.3) works based on the feature selections [5] of particular apps than developer choice. The feature selections of the apps allow and ensure the developer that the permission model implements by its developing platform to automatically generate the required permission model for the particular app or features. So, while developing apps developers does not need to worry about crashing the apps and can use the desire permission [8] as per the selections of the features (i.e. calls, sms, internet etc).

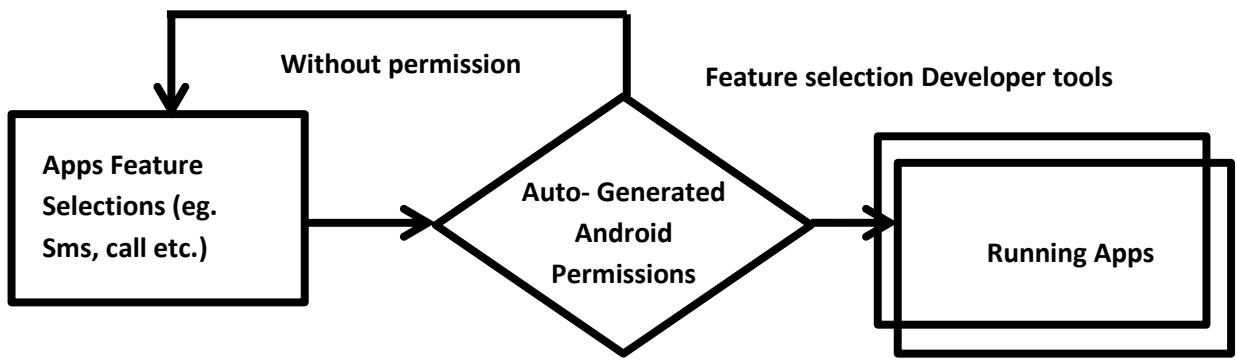

Figure 1.3: Auto Generate Permission Models – Developer platform

Moreover, even if some unnecessary feature use on programming, this model can further implements the manual feature based permission selection by users, which will offer the list of feature and permission for users to select or deny for the app completely or partly. However, apps will not crash but it will disable the denied features by the users. By using this model developer does not need to select the permission, but they can select the feature which is more precise for apps developers, while users are also able to use the function based on their choice without possible security exploit.

*Feature based Manual Permission selections for users:* Including users on decision-making is one of the great development achievements on the smart phone programming, while users are always stay as biggest stakeholders on the smart phone and its apps. The user models support development of the application based on the features [5]. When apps developed as independent features [13] and integrate as apps than creating the bundled apps, the features (like calling, internet, sms etc.) applied as a user decision from the manual selection process when installing apps or while using it.

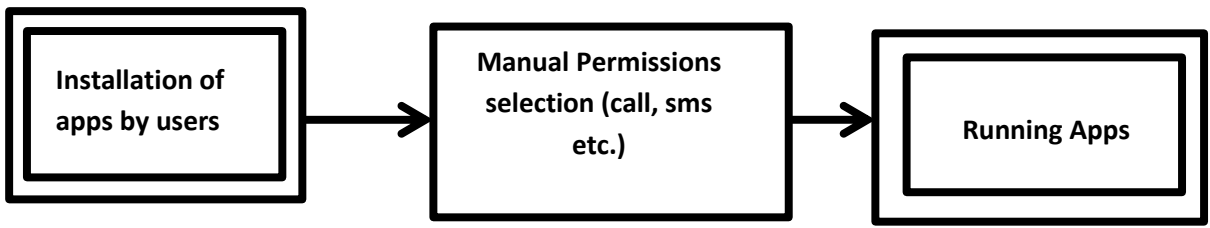

Figure 1.4: Manual Permission Models – For Users

The figure (Figure 1.4) shows the manual selection procedures models which allows users to select the particular and only required permission when installing the apps.

Generally, when users download the apps from play store and try to install on the phone, the apps all allows users to select the individual permission and run it.

*Feature Based Development for Android:* Android has largest market place on smart phones its Google play store filled by malicious apps masquerading as regular apps and steal data from users [12]. The concept behind the feature based development is making apps untouchable by the hackers or malicious apps. So here I present development of the android application as non-executable feature files bundled as applications. Here I purpose development of the features independently rather developing as an executable applications. Developers can develop particular features which will combine in the application bundled and features executed as a standard application files for all the apps [5].

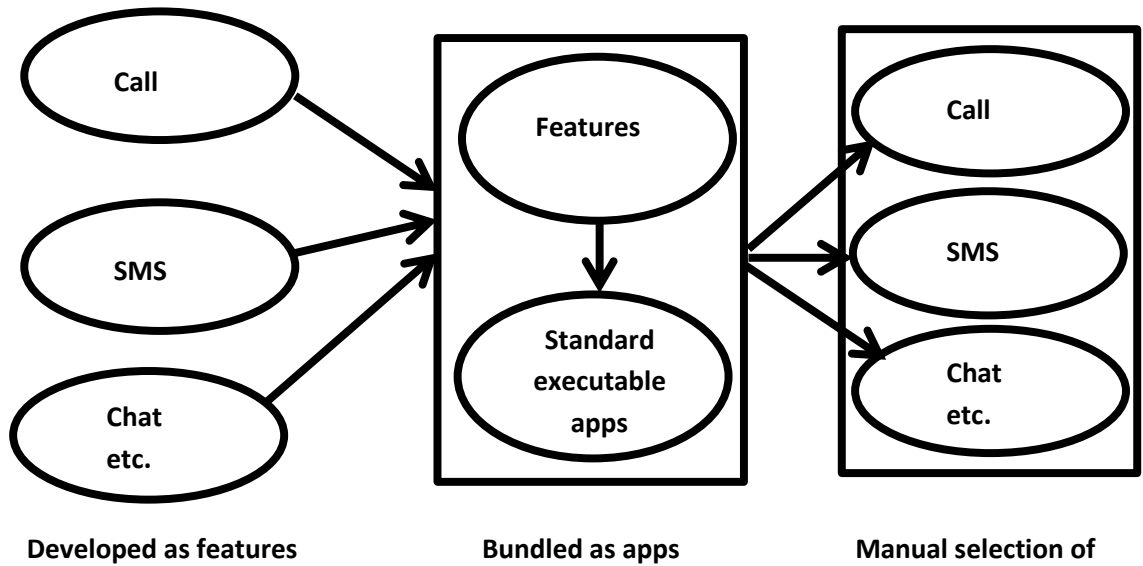

Figure1.5: Feature based development for android

Above figure (Figure 1.5: Feature based development for android) shows the development process of android apps as a feature based development. The process has following main functionalities:

Developing Features: The concept here is development of the features [5] i.e. chat, call, sms etc. as an independent models of the programming rather the application's part.

Executing Features by Standard apps: These feature executed by the standard apps provided by the android or developers which will provided the services as a bundle of the features as an app.

Manual selection of features for users: It allows the storing of the features [2] on the users phones via apps and manual selection of permission (SMS, call, internet etc.) as each individual features.

Overall, feature based development models allows developers to use auto generated permission and developed as features which is bundled and published into market. The users download published apps and select permission for their required features while installing it.

**Future work**
In order to make development and uses more secure on phones this concept is first presented in this paper. Moreover, feature based programming extends the concept of using independent developments of feature of particular apps which will further merged with the main apps activity.

**Conclusions**
Overall, information compromised in different ways by malicious apps and hackers by using weak permission model of android, badly saved or send file in external storage, nonsensical information lies on the source code of application or by using third-party update and rooting mechanism. Different apps on android smart phones bypass the android permission model to have root access. User's personal data can compromise by developers or stolen by its apps.

The structure of android permission model is not enough to stops it. Moreover, android update mechanism works on pushed out so it is possible for third-party to bypass and gain root access. Malware works to bypass the android permission models by using piggy bagged poorly coded apps and stole the user's personal information. However, whose personal information compromise they really don't have much of choice.

Finally, developer and vendor have to work and offer security on user's personal data. So, the solution provided as feature based development programming models implementing feature based selections of the permission allows developers easy way of programming without crashed apps, including user as a key stakeholder for decision-making and development based on the features of particular application so we can provide the sense of security while using the phone apps.

**REFERENCES**
[1] System Permissions, https://developer.android.com/training/permissions/index.html
[2] List of Features on Android, https:// en.wikipedia.org/ wiki/List_ of_features_in_Android,
 Android Version features, www.android.com/versions/
[3] Android Developer/ Developer workflow, developer.android.com/tools/workflow/index.htm
[4] Introduction to Android, http://wideskills.com/android/overview-android/android-definition, Android architecture, http://wideskills.com/android/architecture-of-android
[5] L Ladha, T Deepa, Feature selection methods and algorithms, International journal of computer science and engineering (IJSCE) ISSN : 0975-3397, Vol. 3 No. 5 May 2011, p1787
http://www.enggjournals.com/ijcse/doc/IJCSE11-03-05-051.pdf
[6] Storage Options- External Storage, developer.android.com/guide/topics/data/data-storage.html
[7]  Dangerous permisssion, https://developer.android.com/guide/topics/security/permissions.html
[8] Jeff six, AppsecDC 2012, an in depth introduction to android permission model, https://www.owasp.org/images/c/ca/ASDC12An_InDepth_Introduction_to_the_Android_Permissions_Modeland_How_to_Secure_MultiComponent_Applications.pdf
[9] Vaibhav Rastogi, Yan Chen, Xuxian Jiang, 2013, DroidChameleon: Evaluating Android Anti-malware against Transformation Attacks, Northwestern University, North Carolina State University,vrastogi@u.northwestern.edu,ychen@northwestern.edu,jiang@cs.ncsu.edu,
https://www.csc.ncsu.edu/faculty/jiang/pubs/ASIACCS13_DroidChameleon.pdf
[10] Android Studio, Developer platform, https://developer.android.com/sdk/index.html

[11] Laurent simon, Ross Androson , Security Analysis of Android Factory Resets, University of Cambridge, http://www.cl.cam.ac.uk/~rja14/Papers/fr_most15.pdf

[12] Tiwari Mohini, Srivastava Ashish Kumar and Gupta Nitesh, Review on android and smartphone security, Research journal of computer and information technology science, Vol. 1(6), 12-19, November (2013), ISSN 2320 – 6527, www.isca.in, www.isca.me

[13] Samsung, Samsung knox, security solutions, https://www.samsungknox.com/en/system/files/whitepaper/files/Samsung_KNOX_Security_Solution_V1_10_0.pdf

[14] LuyiXing, Xiaorui Pan, Rui Wang, Kan Yuan and XiaoFeng Wang, Pileup Flaws: Vulnerabilities in Android update make all android device vulnerable, System Security Lab at Indiana University, Microsoft Research, http://www.secureandroidupdate.org